\documentclass{czjphys}         
\usepackage{epsf}
\begin{document}
\title{Relativistic Solenoids\footnote{Dedicated to Prof. Ji\v{r}\'{\i} Hor\'{a}\v{c}ek on the occasion of his $60^{\mbox{{\tiny th}}}$ birthday.}}  
%
\authori{M. \v{Z}ofka and J. Langer}      \addressi{Institute of Theoretical Physics,  Faculty of Mathematics and Physics, Charles University,\\
  V Hole\v{s}ovi\v{c}k\'{a}ch 2, 180 00 Prague 8, Czech Republic}
\authorii{}     \addressii{}
\authoriii{}    \addressiii{}
\authoriv{}     \addressiv{}
\authorv{}      \addressv{}
\authorvi{}     \addressvi{}
%
\headauthor{M. \v{Z}ofka and J. Langer}            
\headtitle{Relativistic Solenoids}             
\lastevenhead{M. \v{Z}ofka and J. Langer: Relativistic Solenoids} 
\refnum{A}
\daterec{XXX}    
\issuenumber{0}  \year{2001}
\setcounter{page}{1}
\maketitle

\begin{abstract}
We construct a general relativistic analogy of an infinite solenoid, i.e., of an infinite cylinder with zero electric charge and non-zero electric current in the direction tangential to the cylinder and perpendicular to its axis. We further show that the solution has a good weak-field limit.
\end{abstract}

\section{Introduction}
In classical electrodynamics, a solenoid is an infinite coil conducting electric current that acts as the source of the homogeneous magnetic field inside (there is no field outside). As a model, we consider an infinite cylinder with vanishing charge but non-zero surface current density. In general relativity, there is no solution involving only a homogenous magnetic field. The only static solution of Einstein-Maxwell equations representing a magnetic field that is cylindrically symmetric and regular on the symmetry axis is the Bonnor-Melvin universe \cite{Melvin} the invariant of which decreases with increasing distance from the axis. This is a natural consequence of the fact that the energy of the magnetic field curves the spacetime. However, a weak field changes only slowly near the axis thus representing a good approximation of the classical situation.

Therefore, we take the Bonnor-Melvin spacetime and join it to the Levi-Civita spacetime, i.e., to the most general static, cylindrically symmetric solution of vacuum Einstein equations. This is achieved in such a way that junction conditions are satisfied on the hypersurface that separates the two spacetimes and represents a thin cylindrical shell of matter. To determine the surface energy-momentum tensor and the electric current on the shell, we apply Israel formalism \cite{Israel} generalized by Kucha\v{r} \cite{Kuchar} to Einstein-Maxwell fields. There is one free parameter in the interior Bonnor-Melvin solution and two free parameters in the exterior Levi-Civita solution. We show that for any circumference of the cylinder we can always choose these three parameters in such a way that the the density and pressure of the matter on the shell satisfy energy conditions. Moreover, the energy-momentum tensor can be interpreted as that of counter-rotating, charged particles held in equilibrium by the Lorentz force and gravity.

With a weak magnetic field inside, low unit-length mass of the shell, and a small radius of the shell, the resulting field corresponds very well to the Newtonian case and classical electrodynamics. However, the picture is substantially different for stronger magnetic and gravitational fields and/or greater radii of the shell.\\
\section{Construction of the Solenoid}
The spacetime metric inside the shell is the Bonnor-Melvin universe \cite{Horsky}
\begin{equation} \label{Bonnor Melvin}
g_{\mu \nu} = \alpha^{-2}(-c^2 \mbox{d}\tau^2+\mbox{d}\zeta^2) + \alpha^{-5}\mbox{d}\rho^2 + \alpha \rho^2 \mbox{d}\varphi^2,
\end{equation}
where
\begin{equation} \label{Definition of alpha}
\alpha = 1 - \frac {G K^2} {c^4} \rho^2,
\end{equation}
and $\rho \leq r < c^2/K \sqrt{G}$.
Constant $K$ determines the strength of the magnetic field since the 4-potential reads
\begin{equation} \label{4-potential}
A = K \rho^2 d\varphi.
\end{equation}
The Maxwell tensor only has one non-zero component
\begin{equation} \label{Maxwell tensor}
F_{\rho \varphi} = 2 K \rho,
\end{equation}
and its invariant reads
\begin{equation} \label{Maxwell field invariant}
F_{\mu \nu} F^{\mu \nu} = 8 K^2 \alpha^4.
\end{equation}
The analogy between the Bonnor-Melvin magnetic field and a special-relativistic homogeneous magnetic field $H$ oriented along the $z$-direction ($F_{\rho \varphi} = H \rho$ and $F_{\mu \nu} F^{\mu \nu} = 2 H^2$) is $H=2K$. However, we must bear in mind that the invariant of the field is not constant in our case---this is to be expected since the electromagnetic field adds to the gravitational mass and a homogeneous field is thus not possible.

The usual form of the metric \cite{Melvin}
\begin{equation} \label{Original Bonnor Melvin metric}
g_{\mu \nu} = \beta^2(-c^2 \mbox{d}\tau^2+\mbox{d}\zeta^2 + \mbox{d}\tilde{\rho}^2) + \tilde{\rho}^2 /\beta^2 \mbox{d}\varphi^2,
\end{equation}
where $\beta = 1 + G K^2 \tilde{\rho}^2 / c^4$ is obtained from (\ref{Bonnor Melvin}) by the transformation $\tilde{\rho}^2 /\beta^2 = \alpha \rho^2$.

The singular surface located at $\rho = c^2/K \sqrt{G}$ at infinite proper distance from the axis cannot be reached in finite time even by photons and all timelike geodesics have their turning points closer to the axis.

\begin{figure}[h]
\begin{center}
\epsfxsize=8cm
\epsfbox{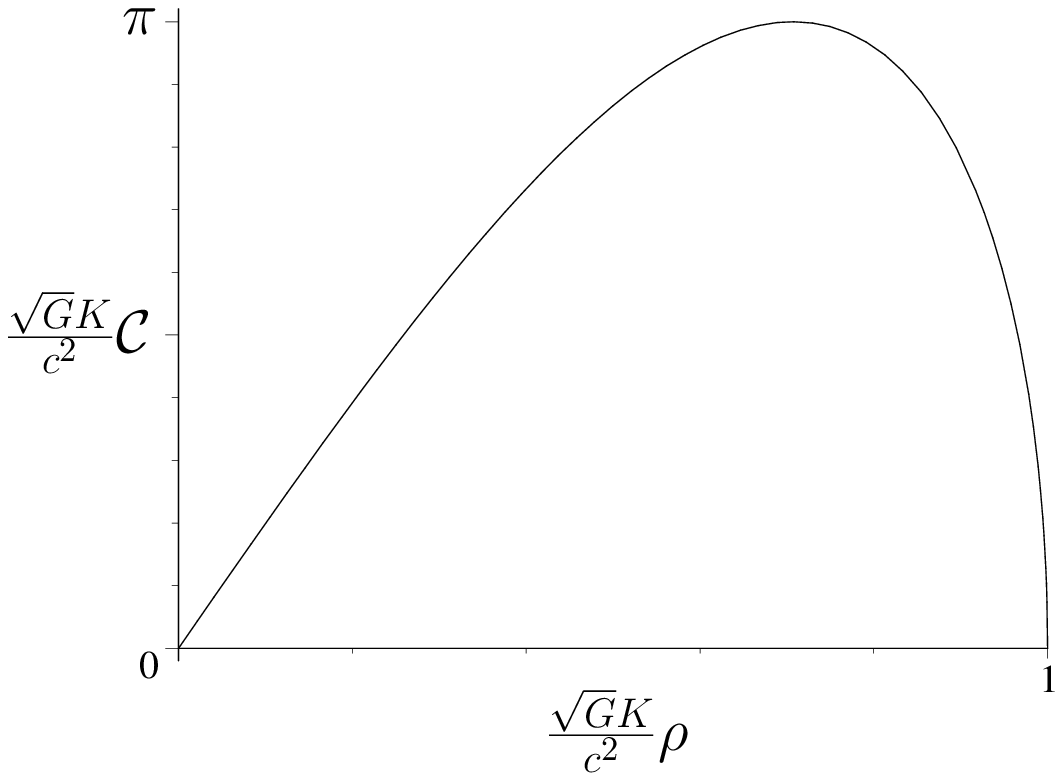}
\end{center}
\caption{\label{Circumference}Circumference $\mathcal{C}$ of rings $\tau, \zeta, \rho =$ constant in Bonnor-Melvin universe as a function of their radius $\rho$. The maximum value is attained at $\rho = c^2 /K \sqrt{2G}$ or $\alpha = 1/2$.}
\end{figure}

The circumference of rings $\tau, \zeta, \rho =$ constant reaches its maximum value of
\begin{equation} \label{Maximum circumference}
\mathcal{C}_{\mbox{\tiny{max}}} = \pi c^2 / K \sqrt{G}
\end{equation}
for $\rho = c^2 /K \sqrt{2G}$ or $\alpha = 1/2$ and then decreases to zero again (see Figure \ref{Circumference}).

The spacetime metric outside the shell is the Levi-Civita universe
\begin{equation} \label{Levi-Civita}
g_{\mu \nu} = -\rho^{2m} c^2 \mbox{d}\tau^2 + \rho^{2m(m-1)} (\mbox{d}\zeta^2 + \mbox{d}\rho^2) + \rho^{2(1-m)}/C^2 \mbox{d}\varphi^2,
\end{equation}
where $C$ is the conicity parameter of the metric, $m$ is the Levi-Civita parameter related to the strength of the gravitational field, and $\rho \geq R$.

We identify both parts of the spacetime along a cylindrical shell of radius $r$ as measured inside and $R$ as measured outside. The only condition we must satisfy is that the length of the hoops around the cylinder be the same from inside and outside
\begin{equation} \label{Hoop constraint}
\sqrt{\alpha}r = R^{(1-m)}/C,
\end{equation}
which we can understand as an equation fixing the conicity parameter $C$ outside. This means that for a given outer spacetime and a fixed value of the magnetic field inside there may be two different inner spacetimes of the same circumference (if this value is less than $\mathcal{C}_{\mbox{\tiny{max}}}$ of (\ref{Maximum circumference})), there could be just one (for this critical value), or none at all---it is not possible to join such an external spacetime to a Bonnor-Melvin universe. We choose the intrinsic coordinates $T, Z, \mathit{\Phi}$ on the shell in such a way that the induced metric is Minkowskian, $T$ being the proper time of static observers and $Z, \mathit{\Phi}$ the proper distances in the $Z$ and $\mathit{\Phi}$ directions.
\section{Interpretation}
We now calculate the properties of the material within the seam using the Israel formalism \cite{Israel}. The resulting surface energy-momentum tensor can be understood in terms of phenomenological pressures or tensions satisfying various energy conditions or it can be given a more direct physical meaning via particle interpretation as shown below. We find
\begin{equation} \label{Induced energy momentum tensor}
\begin{array}{rcl}
8\pi G S_{TT} / c^4 & = & \alpha^{3/2}/r - (1-m)^2/R^{\mathcal{A}}, \\
8\pi G S_{ZZ} / c^4 & = & -\alpha^{3/2}/r + 1/R^{\mathcal{A}}, \\
8\pi G S_{\mathit{\Phi}\mathit{\Phi}} / c^4 & = & -4 (1-\alpha) \alpha^{3/2} / r + m^2/R^{\mathcal{A}},
\end{array}
\end{equation}
where $\mathcal{A} = m^2-m+1$ and $\alpha$ is evaluated at $\rho=r$ and is thus a function of $r$ and $K$. The induced 3-current has only the axial component since $J_a = [F_{a\bot}] \equiv (F_{a\bot})_{\tiny\mbox{out}} - (F_{a\bot})_{\tiny\mbox{in}} = -(F_{a\bot})_{\tiny\mbox{in}}$:
\begin{equation} \label{Induced 3-current}
J_\mathit{\Phi} = \frac {c} {2 \pi} K \alpha^2.
\end{equation}
It is to be noted that this value only depends on the inner spacetime parameters. If we increase the inner radius of the cylinder up to its maximum value then $\alpha \rightarrow 0$, the circumference of the `solenoid' decreases, and so does the surface current for any intensity of the magnetic field on the axis. Therefore, it is difficult to understand the current as the source of the magnetic field as in the classical case.

The surface energy-momentum tensor (\ref{Induced energy momentum tensor}) can be interpreted as due to four streams of charged particles spiraling along the axis with
\begin{equation} \label{4 streams - velocities}
\begin{array}{rclrcl}
U_1 & = & \gamma [c, v_Z, v_\mathit{\Phi}], & U_2 & = & \gamma [c, v_Z, -v_\mathit{\Phi}], \\
U_3 & = & \gamma [c, -v_Z, v_\mathit{\Phi}], & U_4 & = & \gamma [c, -v_Z, -v_\mathit{\Phi}],
\end{array}
\end{equation}
where $\gamma=1 / \sqrt{ 1-v_Z^2/c^2-v_\mathit{\Phi}^2/c^2}$. The corresponding energy-momentum tensor $S_{ij} = \rho ( U_{1i} U_{1j} + U_{2i} U_{2j} + U_{3i} U_{3j} + U_{4i} U_{4j})$, where $\rho$ is the surface rest density of the rest mass of the streams, has the following form
\begin{equation} \label{4 streams - energy momentum tensor}
S_{ij} = 4\rho \gamma^2 \left(
\begin{array}{ccc}
c^2 & 0 & 0 \\
0 & v_Z^2 & 0 \\
0 & 0 & v_\mathit{\Phi}^2\\
\end{array}
\right).
\end{equation}
For the 3-current $J = \sigma (U_1 - U_2 + U_3 - U_4)$, with $\sigma$ being the surface rest density of the charge of the streams, we obtain
\begin{equation} \label{4 streams - 3-current}
J_i = 4 \sigma v_\mathit{\Phi} \gamma [ 0, 0, 1 ].
\end{equation}
Comparing these expressions to the induced energy-momentum tensor (\ref{Induced energy momentum tensor}), we derive
\begin{equation} \label{Properties of the 4 streams}
\begin{array} {rclrcl}
\rho & = & \frac {c^2}{16 \pi G} \left( \frac {\alpha^{3/2}} {r} (3-2\alpha) - \frac {\mathcal{A}} {R^{\mathcal{A}}}\right), & \sigma^2 & = & \frac {K^2 \alpha^4} {64 \pi^2} { \frac {\alpha^{3/2} (3-2\alpha) /r - \mathcal{A}/R^{\mathcal{A}}} {m^2/R^{\mathcal{A}} - 4 \alpha^{3/2} (1-\alpha) /r}},  \vspace{0.1cm}
\\
v_Z^2 & = & c^2 \frac {1/R^{\mathcal{A}} - \alpha^{3/2}/r} {\alpha^{3/2}/r - (1-m)^2/R^{\mathcal{A}}}, & v_\mathit{\Phi}^2 & = & c^2 \; \frac {m^2/R^{\mathcal{A}} - 4 \alpha^{3/2} (1-\alpha) /r} {\alpha^{3/2}/r - (1-m)^2/R^{\mathcal{A}}}.
\end{array}
\end{equation}
Another physical characteristic of the cylinders is their mass per unit length of the shell
\begin{equation}\label{M1 - definition}
M_1 \equiv \mbox{(Circumference)} \cdot \; S_{TT} = 2\pi r \alpha^{1/2} S_{TT} = \frac {c^2} {4G} \left[ \alpha^2 - \frac {(1-m)^2} {C R^{m^2}} \right].
\end{equation}
It should be noted that its value is always less than $1/4$---the same restriction holds also for solid cylinders \cite {Anderson}, \cite{BLZS} and in the absence of magnetic fields \cite{BZ}. 

\begin{figure}[h]
\begin{center}
\epsfxsize=8cm
\epsfbox{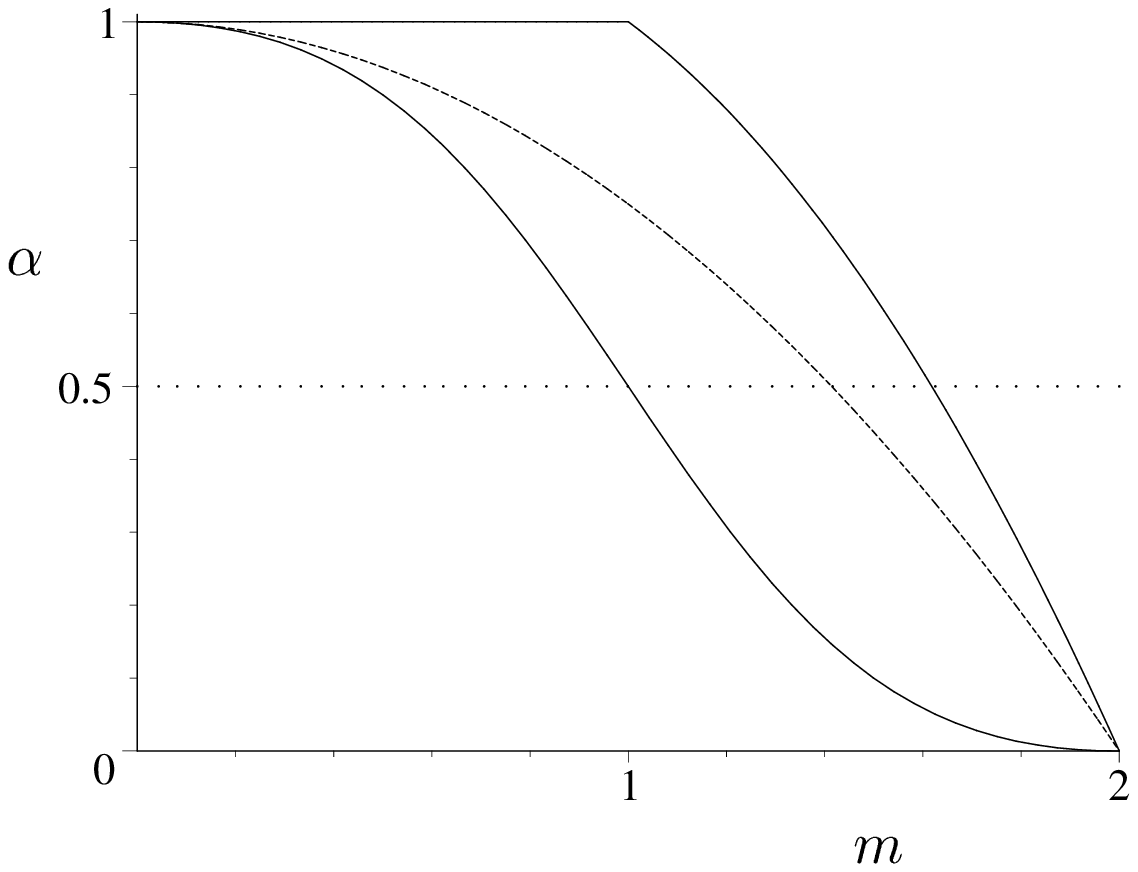}
\end{center}
\caption{\label{Energy Conditions} Interpretation ranges for counter-rotating and counter-spiraling particles: $\alpha$ corresponds to the inner radius of the cylinders and $m$ is the outer Levi-Civita parameter. Counter-spiraling particles admit solutions between the two full curves while counter-rotating particles only between the upper full curve and the dotted curve. In these areas we can always find such a combination of the 5 spacetime parameters that the resulting shell can be made of particles. Solutions along the $\alpha=0.5$ line reach the maximum possible circumference.}
\end{figure}

Interpretation can become considerably simpler if the $S_{ZZ}$ component vanishes. We thus require
\begin{equation} \label{Azimuthal motion only}
R^{\mathcal{A}} = r/\alpha^{3/2},
\end{equation}
obtaining
\begin{equation} \label{2 streams - induced energy momentum tensor}
S_{ij} = \frac {\alpha^{3/2} c^4} {8 \pi G r} \left(
\begin{array}{ccc}
m(2-m) & 0 & 0 \\
0 & 0 & 0 \\
0 & 0 & m^2 - 4(1-\alpha)\\
\end{array}
\right).
\end{equation}
In this case it is sufficient to only consider two oppositely charged streams of particles moving in the azimuthal direction with $U_1 = \gamma [c, 0, v_\mathit{\Phi}], U_2 = \gamma [c, 0, -v_\mathit{\Phi}]$ with $\gamma = 1/\sqrt{1-v_\mathit{\Phi}^2/c^2}$. Therefore
\begin{equation} \label{2 streams - energy momentum tensor}
S_{ij} = 2\rho \gamma^2 \left(
\begin{array}{ccc}
c^2 & 0 & 0 \\
0 & 0 & 0 \\
0 & 0 & v_\mathit{\Phi}^2\\
\end{array}
\right).
\end{equation}
For the 3-current $J_i = \sigma (U_1 - U_2)$, we obtain
\begin{equation} \label{2 streams - 3-current}
J_i = 2 \sigma v_\mathit{\Phi} \gamma [ 0, 0, 1 ].
\end{equation}
Comparing these expressions to expression (\ref{2 streams - induced energy momentum tensor}), we find
\begin{equation} \label{Properties of the 2 streams}
\begin{array} {rcl}
\rho & = & \frac {\alpha^{3/2} c^2}{8 \pi G r} \left( (3-2\alpha) - \mathcal{A}\right), \nonumber \vspace{0.1cm}\\
v_\mathit{\Phi}^2 & = & c^2 \frac {m^2 - 4 (1-\alpha)} {m(2-m)}, \nonumber  \vspace{0.1cm}\\
\sigma^2 & = & \frac {K^2 \alpha^4} {8 \pi^2} { \frac {(3-2\alpha) - \mathcal{A}} {m^2 - 4 (1-\alpha)}}.
\end{array}
\end{equation}
Finally, we find
\begin{equation}\label{M1 - 2 streams}
M_1 = \frac {\alpha^2 c^2} {4G} m(2-m).
\end{equation}
For the ratio of the charge rest density of the streams to their rest mass rest density, we obtain
\begin{equation}\label{epsilon - 2 streams}
\varepsilon^2 \equiv \left( \frac{\sigma}{\rho} \right)^2 = 8G \frac { \alpha ( 1-\alpha )} {[m^2 - 4(1-\alpha) ] [m(1-m) + 2(1-\alpha)]}.
\end{equation}
The admissible range of $\alpha$ for a given $m$ is given by the requirements $\rho, v_\mathit{\Phi}^2, \sigma^2 >0$ and $v_\mathit{\Phi}^2 <1$. This gives $m \in (0 ; 2)$ and $\alpha \in ( 1-m^2/4; 1 - m^2/2 + m/2)$ as can be seen in Figure \ref{Energy Conditions}. All these shells can be constructed of counterrotating particles moving in the azimuthal direction only. For $m \in ( \sqrt{2} ; ( 1 + \sqrt{5} )/2 )$ the shells can reach their maximum possible circumference. If we fix all three parameters of the outer spacetime ($m, R, C$) then there is only a single inner spacetime ($K, r$) that can be attached to it due to the conditions (\ref{Hoop constraint}) and (\ref{Azimuthal motion only}).

If we go back to the more general case of the spiraling particles, the allowed range of the parameters is broader, $m \in (0 ; 2)$ and $\alpha \in ( (2-m)^2/ 2 (m^2-2m+2); 1 - m^2/2 + m/2)$, and includes the above case of the azimuthal particles (Figure \ref{Energy Conditions}). In this case it is possible to completely specify the outer spacetime and still find different inner solutions (however, the corresponding induced energy-momentum tensors will be different). Therefore, a given Levi-Civita spacetime can be thought of as admitting diverse possible sources. (This is due to the fact that the junction conditions on the shell surface do not require the continuity of the first derivatives of the characteristic forms of the spacetime.) We should also note that if we want to reach the maximum possible circumference of the shell ($\alpha=1/2$) we must admit $m \ge 1$ entering thus the region where the Levi-Civita spacetime does not have a classical counterpart.

It is interesting that if we insist that there be Minkowski spacetime inside the cylinder \cite{BZ} then the particle (or photon) interpretation only allows $m \in [0 ; 1]$. By admitting magnetic fields inside the cylinder, we extended the particle-inter\-pretation range for the sources of the Levi-Civita spacetime. Curiously, if we investigate the weak, strong, and dominant energy conditions, the range of $m$ remains unchanged.
\section{The Weak-field Limit}
First we summarize the results for classical cylinders composed of two  streams of oppositely charged particles moving in $+\mathit{\Phi}$ and $-\mathit{\Phi}$ directions (if we admit axial current we find there is a nonzero magnetic field outside the cylinder which we do not consider the classical counterpart of the relativistic cylinders treated above). The properties of the particles are characterized by the ratio of their charge to their rest mass, $\varepsilon$, and their density is described by the mass per unit length of the cylinder, $M_1$. The azimuthal velocity of the streams is
\begin{equation} \label{Classical cylinders - azimuthal velocity}
v_\mathit{\Phi} = \sqrt{\frac {G} {\varepsilon^2/c^2 + 1/M_1}}.
\end{equation}
Expressing $\varepsilon$ as a function of the remaining parameters we obtain
\begin{equation} \label{Classical cylinders - epsilon}
\varepsilon = c \sqrt{G/v_\mathit{\Phi}^2 - 1/M_1}.
\end{equation}
If we have a cylinder of radius $r$ made of charged fluid of pressure $p$ ($p = \rho v_{\mathit{\Phi}}^2$ due to the streaming of the particles composing the fluid) that induces magnetic field $H$ inside, we find
\begin{equation} \label{Classical fluid}
G M_1^2 = 2 \pi r p + \frac {1} {4} r^2 H^2 \equiv P + \frac {1} {4} r^2 H^2,
\end{equation}
where $P = 2 \pi r p = \mathcal{C} p$ is pressure integrated around the cylinder.

We can now compare the classical cylinders to the weak-field limit of the relativistic results. We base our interpretation on one of the junction conditions due to Kucha\v{r} \cite{Kuchar}, namely
\begin{equation} \label{Junction condition}
\tilde{K}_{AB} S^{AB} = \frac{1}{c} \tilde{F}_{A\bot}J^A,
\end{equation}
where $\tilde{K}_{AB} = (K^+_{AB} + K^-_{AB})/2$ is the mean extrinsic curvature of the surface, $S_{AB}$ is the induced energy-momentum tensor, $\tilde{F}_{A\bot} = (F^+_{A\bot} + F^-_{A\bot})/2$ is the mean Maxwell tensor, $J_A$ is the induced current, and upper-case Latin indices take on values $T, Z, \mathit{\Phi}$. In fact, (\ref{Junction condition}) is not an additional constraint on the solution---it is automatically satisfied here since both spacetimes obey the Einstein-Maxwell equations. We now interpret $S_{TT}$ as the energy density, $\rho c^2$, of the induced matter and $S_{\mathit{\Phi}\mathit{\Phi}}$ as its pressure, $p$. We find
\begin{equation} \label{Balance equation}
- \frac {1} {2r} \alpha^{3/2} (m + 2(1-\alpha)) \rho c^2 - \frac {1} {2r} \alpha^{3/2} (m-2\alpha) p = - \frac {1} {2 \pi} K^2 \alpha^4.
\end{equation}
We first set $K=0$ and thus $\alpha=1$ to check the uncharged case (Minkowski inside). We obtain
\begin{equation}
m\rho + (m-2) p = 0.
\end{equation}
In the weak-field limit of $m \rightarrow 2 G M_1 / c^2 \rightarrow 0$ (see (\ref{M1 - definition}) and also \cite{BZ}) and $M_1 = \mathcal{C} \rho = 2 \pi r \rho$, where $\mathcal{C}$ is the shell circumference, we finally have the correct formula (see (\ref{Classical fluid}))
\begin{equation}
G M_1^2 - P = 0,
\end{equation}
where $P = \mathcal{C} p$. If we keep the Lorentz force in our equation, we have
\begin{equation}
M_1 c^2 (m + 2(1-\alpha)) + P (m-2\alpha) = 2 r^2 K^2 \alpha^3,
\end{equation}
with $\mathcal{C} = 2 \pi r \sqrt{\alpha}$. If we are interested in cylinders of small radii in which the electromagnetic field does not contribute much to their gravitational mass, then $\alpha \rightarrow 1$ and we obtain
\begin{equation}\label{Approximate balance equation}
M_1 c^2 (m + 2(1-\alpha)) - 2 P = 2 r^2 K^2.
\end{equation}
Assuming further $2(1-\alpha) \ll m \ll 1$ and remembering that $H=2K$, we finally find
\begin{equation}
G M_1^2 =  P + \frac{1}{4} r^2 H^2,
\end{equation}
which is the same as (\ref{Classical fluid}). To keep higher-order terms in the equation, we rewrite (\ref{Approximate balance equation}) as follows:
\begin{equation}\label{Balance with K}
2 G M_1 (\frac{M_1}{2} + \frac {K^2 r^2} {2c^2}) - P = r^2 K^2.
\end{equation}
If we integrate $F_{\mu\nu} F^{\mu \nu}/16 \pi$ (the energy density of the electromagnetic field) within the cylinder up to unit proper length in $z$, we find
\begin{equation}\label{Electromagnetic Energy}
E_1 = \int_0^{2\pi} \!\!\!\!\! d\varphi \int_0^r \!\! d\rho \int_0^{\alpha} \!\!\! dz \: \frac{\rho}{\alpha^3} \: \frac{K^2 \alpha^4}{2 \pi} = \frac{1}{2} (K^2 r^2 - \frac{G}{c^4} K^4 r^4 + \frac{G^2}{3 c^8} K^6 r^6).
\end{equation}
Bearing in mind that the mass of the shell $M_1$ acts on itself with a weight factor of 1/2, we see that the lowest-order term in (\ref{Electromagnetic Energy}) corresponds exactly to the additional term in the gravitational mass in (\ref{Balance with K}). This is a nice confirmation of the fact that the energy of the electromagnetic field contributes to the gravitating mass of the cylinder.

A direct comparison between the general relativistic formula for the ratio of the charge rest density of the streams to their rest mass rest density (\ref{epsilon - 2 streams}) and its classical counterpart (\ref{Classical cylinders - epsilon}) is not possible since (\ref{epsilon - 2 streams}) is expressed in terms of $\alpha$ and $m$. Although a straightforward substitution for these parameters in terms of $M_1$ and $v_\mathit{\Phi}$ from (\ref{Properties of the 2 streams}) and (\ref{M1 - 2 streams}) does not yield a simple algebraic formula, a series expansion verifies their correspondence.
\section*{Acknowledgements}
We thank Tom\'a\v{s} Ledvinka for discussions. We were supported in part by Grants Nos. GACR 202/02/0735 and GAUK 141/2000 of the Czech Republic and the Charles University.


\begin{thebibliography}{15}
    \bibitem {Melvin} Bonnor W B 1954 {\it Proc. Phys. Soc. Lond.} A 67, 225\\
Melvin M A 1964 {\it Phys. Lett.} 8, 65
    \bibitem {Israel} Israel W 1966 {\it Nuovo Cimento} {\bf  B 44} 1 (erratum {\bf B 49} 463)
    \bibitem {Kuchar} Kucha\v{r} K 1968 {\it Czech. J. Phys. B} {\bf 18} 435
    \bibitem {Horsky} Klep\'{a}\v{c} P and Horsk\'{y} J 2001 {\it Czech. J. Phys.} {\bf 51} 1178 and gr-qc/0202027
    \bibitem {Anderson} Anderson M R 1999 {\it Class. Quantum Grav.} {\bf 16} 2845    
    \bibitem {BLZS} Bi\v{c}\'ak J, Ledvinka T, \v{Z}ofka M, and Schmidt B G 2004 {\it Class. Quantum Grav.} {\bf 19} 3653    
    \bibitem {BZ} Bi\v{c}\'ak J and \v{Z}ofka M 2002 {\it Class. Quantum Grav.} {\bf 19} 3653
\end{thebibliography}
\end{document}